# Coexistence of ferromagnetism, antiferromagnetism, and superconductivity in magnetically anisotropic (Eu,La)FeAs$_2$

Jia Yu,[1*] Congcong Le,[2] Zhiwei Li,[3] Lisi Li,[1] Tong Liu,[4,5] Zengjia Liu,[1] Bo Zhang,[3] Bing Shen,[1] Binbin Ruan,[4,5] Zhi'an Ren,[4,5*] Meng Wang[1*]

[1] Center for Neutron Science and Technology, School of Physics, Sun Yat-sen University, Guangzhou 510275, China

[2] Max Planck Institute for Chemical Physics of Solids, 01187 Dresden, Germany

[3] Key Lab for Magnetism and Magnetic Materials of the Ministry of Education, Lanzhou University, Lanzhou 730000, China

[4] Beijing National Laboratory for Condensed Matter Physics, Institute of Physics, Chinese Academy of Sciences, Beijing 100190, China

[5] School of Physical Sciences, University of Chinese Academy of Sciences, Beijing 100049, China

[*] Corresponding authors. E-mails: jyu_work@163.com; renzhian@iphy.ac.cn; wangmeng5@mail.sysu.edu.cn.

**ABSTRACT**

Materials with exceptional magnetism and superconductivity usually conceive emergent physical phenomena. Here, we investigate the physical properties of the (Eu,La)FeAs$_2$ system with double magnetic sublattices. The parent EuFeAs$_2$ shows anisotropy-associated magnetic behaviors, such as Eu-related moment canting and exchange bias. Through La doping, the magnetic anisotropy is enhanced with ferromagnetism of Eu$^{2+}$ realized in the overdoped region, and a special exchange bias of the superposed ferromagnetic/superconducting loop revealed in Eu$_{0.8}$La$_{0.2}$FeAs$_2$. Meanwhile, the Fe-related antiferromagnetism shows unusual robustness against La doping. Theoretical calculation and $^{57}$Fe Mössbauer spectroscopy investigation reveal a doping-tunable dual itinerant/localized nature of the Fe-related antiferromagnetism. Coexistence of the Eu-related ferromagnetism, Fe-related robust antiferromagnetism, and superconductivity is further revealed in Eu$_{0.8}$La$_{0.2}$FeAs$_2$, providing a platform for further exploration of potential applications and emergent physics. Finally, an electronic phase diagram is established for (Eu,La)FeAs$_2$ with the whole superconducting dome adjacent to the Fe-related antiferromagnetic phase, which is of benefit for seeking underlying clues to high-temperature superconductivity.

**INTRODUCTION**

Magnetism is believed to play an important role in high-temperature superconducting pairing, e.g., the Fe-related antiferromagnetism (Fe-AFM) in the iron-based superconducting family[1-3]. The competition between superconductivity (SC) and Fe-AFM in charge-lightly-doping region has been widely revealed[4-8], however, the systems with unusual phase diagrams are also worth our concern. Typically, in the 112-type (Ca,La)FeAs$_2$[9], the Fe-AFM exhibits robustness and is abnormally enhanced by La doping in the overdoped region, with SC gradually suppressed[10]. Lately, a series of the homogenous (Eu,La)FeAs$_2$ compounds were discovered[11]. The transport and magnetic measurements suggested a structural transition (110 K), a Fe-related antiferromagnetic (Fe-AF) transition (98 K), and a Eu-AF transition (46 K) for single-crystalline EuFeAs$_2$[12]. A recent Mössbauer spectroscopy investigation on the polycrystalline EuFeAs$_2$ sample confirmed an incommensurate spin-density-wave-type (SDW-type) AFM ordering of Fe$^{2+}$ around 106 K[13]. The transport measurements on the underdoped Eu$_{1-x}$La$_x$FeAs$_2$ ($x$ = 0–0.15) suggest that the Fe-AFM exists in the studied doping region[11]. Hence, the unusual relationship between Fe-AFM and SC in (Eu,La)FeAs$_2$ is anticipated in a broader doping region, than that of (Ca,La)FeAs$_2$ with lightly-doped samples unavailable.

On the other hand, various Eu-related magnetic properties were revealed in polycrystalline EuFeAs$_2$ under a low magnetic field of 10 Oe, including a spin glass (SG) transition, reentrant magnetic modulation, and moment canting induced by transition metal doping in the Fe site[14,15]. The SG and the moment canting indicate a tunable competition and coexistence of the ferromagnetic and AF interactions between the Eu$^{2+}$ ions, which was proposed to mainly originate in the Ruderman-Kittel-Kasuya-Yosida (RKKY) indirect exchange.

More intriguing is that the coupling between the two magnetic sublattices (see the crystal structure[12] in Fig. 1a) would lead to anisotropic interaction between Eu$^{2+}$ and Fe$^{2+}$ in EuFeAs$_2$. Magnetic systems with anisotropic interactions exhibits various magnetic properties, including sign-reversible exchange bias (EB)[16], spin reorientation (SR)[17], thermal magnetic hysteresis[18], etc. The most studied EB effect is an exchange anisotropy with a shift of the magnetic hysteresis loop along the magnetic-field axis, which was first discovered in oxide-coated cobalt particles with moment compensation in the ferromagnetic/AF interface of Co/CoO[19]. Later, single-phase compounds with double magnetic sublattices have been found to exhibit EB effect due to the existence of anisotropic interactions[20,21]. One explanation is that when a net moment is induced in one of the sublattices by the anisotropic interaction, a circumstance analogous to the ferromagnetism (FM)/AFM interface generates with compensation in between. The moment compensation between FM and SDW-type AFM could also trigger EB effect in alloys or interfaces [22,23]. However, EB effect associated with SDW-type AFM in a stoichiometric compound system is rare. (Eu,La)FeAs$_2$ with robust SDW-type AFM and doping-modifiable Eu-related magnetism is a suitable compound system for exploring the EB anisotropy. Furthermore, the interplays of exotic magnetism and SC have shown interesting physics and application prospects in layered or wire-like heterostructures[24-29]. Hence, the (Eu,La)FeAs$_2$ system is worth deeper investigation, not only for the unusual relationship between SDW and SC, but also for the underlying physics originating from the interplay between anisotropic magnetism and SC.

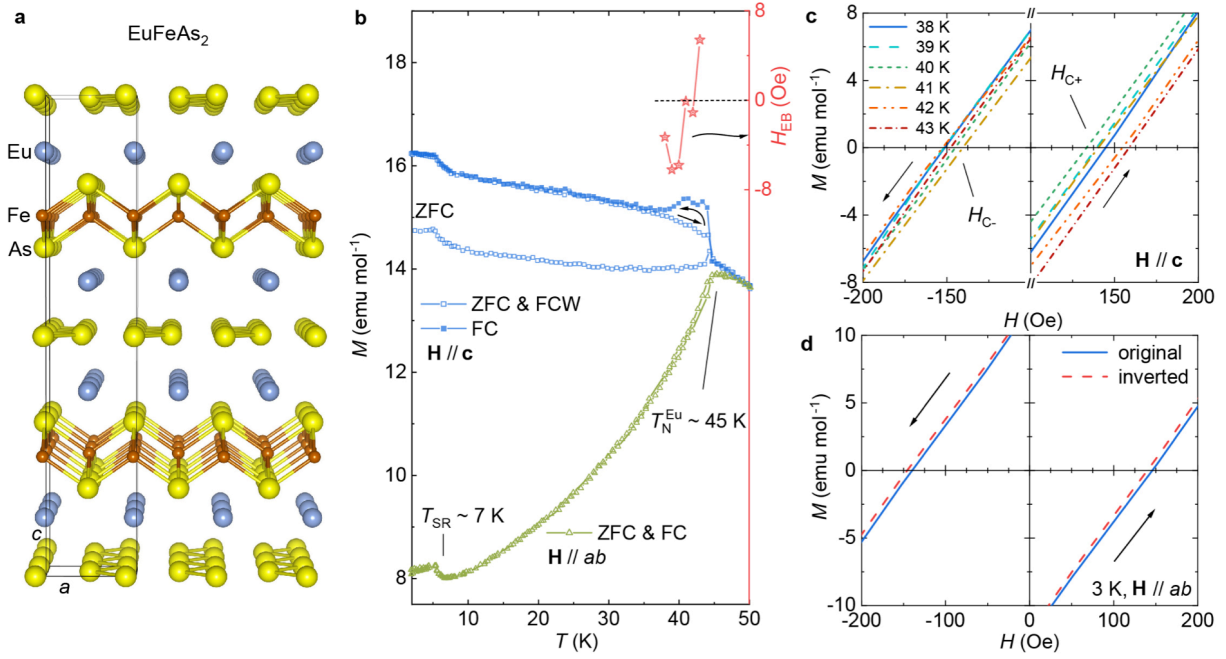

**Fig. 1 Crystal structure and magnetic properties of EuFeAs$_2$. a** Schematic diagram of the crystal structure[12]. **b** Magnetization against temperature under a magnetic field of 100 Oe. **c**, **d** Magnetization against magnetic field at different temperatures with fields parallel to the *c* direction and the *ab* plane, respectively. The stars to the right axis in **b** represent the EB fields of the hysteresis loops in **c**. The original curve in **d**, solid, is the obtained *M-H* curve. The inverted curve, dashes, is the centrosymmetric one of the original.

In this article, we first illuminate the magnetic anisotropy in the parent EuFeAs$_2$. Then, the La-doping-induced magnetic evolution and the coupling between anisotropic magnetism and SC are studied. The nature of the robust Fe-AFM is discussed and examined in the superconducting state. Finally, a La-doping phase diagram on structure, magnetism, and SC is established.

## RESULTS AND DISCUSSION

**Magnetic anisotropy in EuFeAs$_2$.** The phase transitions of EuFeAs$_2$ are reexamined by heat capacity, high-field magnetization, and single-crystal X-ray diffraction (SXRD) analyses, detailed in Supplementary Figure 1. Based on the phase transitions, zero-field-cooling (ZFC), field-cooling (FC), and field-cooled-warming (FCW) magnetization measurements were performed on single-crystalline EuFeAs$_2$ under a low magnetic field of 100 Oe below 50 K. The temperature dependent magnetization (*M-T*) curves, depicted in Fig. 1b, exhibit Eu-related AF moment canceling in the *ab* plane and moment canting in the *c* direction below $T_N^{Eu}$ ~ 45 K. Considering that the RKKY interaction in a conducting system[30] and the anisotropic interaction in an insulating system[31] can both induce net moment, we ascribe the moment canting in EuFeAs$_2$ to the collaboration of the RKKY interaction between Eu$^{2+}$ ions and the anisotropic interaction between Eu$^{2+}$ and Fe$^{2+}$. An SR-like upturn of the magnetization appears below $T_{SR} \sim 7$ K in both directions, corresponding to the reentrant magnetic modulation proposed in the polycrystalline sample[14], while, the SG behavior around 15.5 K is absent in single-crystalline EuFeAs$_2$, even with the magnetic field decreased to 10 Oe, see Supplementary Figure 2. The hysteresis below $T_N^{Eu}$ for the FC and FCW curves in the $c$ direction is reminiscent of the behavior observed in magnetically anisotropic SmCr$_{1-x}$Fe$_x$O$_3$, which is attributed to the lower-temperature SR[32]. Though the temperature interval of the hysteresis in EuFeAs$_2$ is well above $T_{SR}$, the hysteresis still implies a metastable spin state probably fixed by the magnetic anisotropy.

The net moment of Eu$^{2+}$ encourages us to explore the EB anisotropy in undoped EuFeAs$_2$. The magnetization versus magnetic field ($M$-$H$) is studied at different temperatures in the thermal-hysteresis interval with magnetic fields parallel to the $c$ direction, shown in Fig. 1c (full curves presented in Supplementary Figure 2). A magnetic hysteresis behavior appears in the $M$-$H$ curves, with EB discernible from the comparison between the coercivities $H_{C+}$ and $H_{C-}$. The EB fields, defined as

$$H_{EB} = (H_{C+} + H_{C-})/2, \tag{1}$$

are summarized in Fig. 1b. The non-monotonic temperature dependence of $H_{EB}$, is similar to the oscillation behavior associated with incommensurate SDW in the (100)Cr/Ni$_{81}$Fe$_{19}$ bilayers[23]. We ascribe this EB behavior in EuFeAs$_2$ to the anisotropic interaction between Eu$^{2+}$ and Fe$^{2+}$. Besides, a sign reversal of EB occurs in the thermal hysteresis interval, which may be related to the metastable spin state. We also performed a magnetization measurement on EuFeAs$_2$ with **H** // $ab$ at 3 K, as shown in Fig. 1d. Similar to the scenario of **H** // **c**, a weak EB is observed for **H** // $ab$ below $T_{SR}$, as seen from the comparison between the original and inverted curves. Thus, a weak net moment and a moment compensation emerge in the $ab$ plane as well. The FCW measurement below $T_{SR}$ with **H**//$ab$ has also been conducted, while, the magnetization curve basically overlaps with the FC and ZFC curves due to the measurement error of PPMS.

In a word, the single-crystalline EuFeAs$_2$ shows various magnetic properties, mainly associated with the magnetic anisotropy. The EB behaviors related to SDW-type AFM in a stoichiometric compound system enrich the EB effect and the platforms for investigating the mechanism of EB anisotropy.

**La-doping effects in Eu$_{0.79}$La$_{0.21}$FeAs$_2$**. La-doping effects are investigated in overdoped single-crystalline Eu$_{0.79}$La$_{0.21}$FeAs$_2$, of which the doping level is determined by an energy dispersive X-ray spectroscopy (EDXS) analysis, detailed in Supplementary Figure 5. The electrical transport measurement was carried out to check the structural and Fe-AF transitions in this overdoped sample, as demonstrated in Fig. 2a. The resistivity curve exhibits an anomaly around 80 K, similar to that around 100 K in the parent EuFeAs$_2$[12]. According to the phase transitions of the parent EuFeAs$_2$ (see Supplementary Figure 1), the derivation of the $R$-$T$ curve indicates that the structural and Fe-AF transitions are suppressed to $T_S \sim 82$ K and $T_N^{Fe} \sim 73$ K, respectively. The slight resistivity decreasing below 8 K indicates that SC is greatly destroyed, despite the remaining of the robust Fe-AFM.

The $M$-$T$ curves, demonstrated in Fig. 2b and c, exhibit a dramatic ferromagnetic transition at $T_M^{Eu} \sim 32$ K for both directions, and an SR-like transition at $T_{SR} \sim 26.5$ K (determined from the derivation of $M$-$T$ in Fig. 2d) in the $ab$ plane. The magnetic susceptibility can be suppressed by larger fields (not demonstrated), manifesting the canted AF nature of the Eu-FM with field-modifiable competing ferromagnetic and AF interactions. Thermal hysteresis of the FC and FCW processes exists in both directions below $T_M^{Eu}$, which is probably associated with the magnetic anisotropy. No superconducting diamagnetic behavior appears below 8 K.

For comparison, a series of Pr-doped (Eu,Pr)FeAs$_2$ samples are synthesized, of which Pr doping introduces equal electrons but with less magnetic dilution comparing to equally La doping. All the Pr-doped samples exhibit weak moment canting behaviors, even in the overdoped region, detailed in Supplementary Figure 8. Thus, the emergence of the stronger FM in (Eu,La)FeAs$_2$ can be mainly attributed to the La-doping-induced magnetic dilution effect, rather than simply the doping-introduced extra electrons modifying the RKKY interaction. We consider the origin of the net moment in Eu$_{0.79}$La$_{0.21}$FeAs$_2$ same with that in the parent phase discussed above. Then, there are two possible ways to trigger the magnetic dilution effect on enhancing the magnetic anisotropy and generate FM: 1) The anisotropic exchange between Eu$^{2+}$ and Fe$^{2+}$ is adjusted by introducing nonmagnetic La$^{3+}$, which leads to the change of the ferromagnetic-AF competition, similar to the dilution effect in (Sm,La)FeO$_3$[33]; 2) The nonmagnetic La$^{3+}$ will not participate the RKKY interaction, which results in the doubling of the interaction distance between the Eu$^{2+}$ moments beside the La$^{3+}$ ion and might change the proportion of the ferromagnetic term of the RKKY interaction.

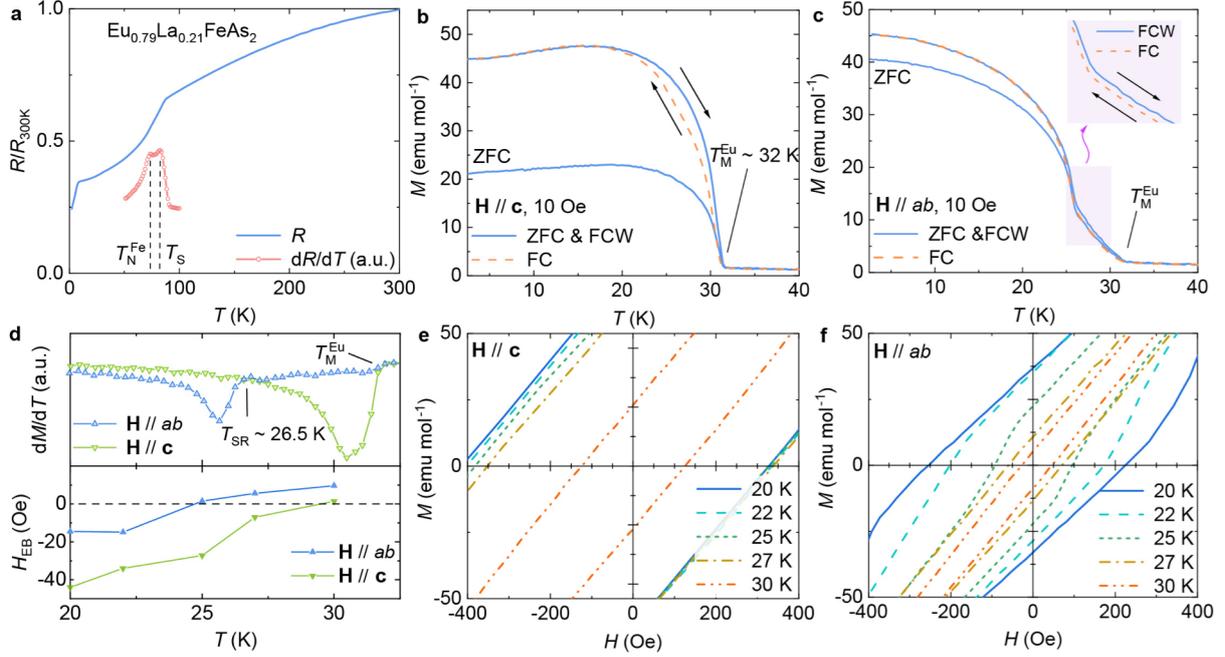

**Fig. 2 Physical properties of Eu$_{0.79}$La$_{0.21}$FeAs$_2$. a** Normalized in-plane resistivity against temperature with part of the first derivation curve. **b, c** Magnetization against temperature with magnetic fields parallel to different directions. **d** First derivation of the *M-T* curves (10 Oe) and the EB fields in different directions. **e, f** Magnetization against field in different directions. The inset in **c** is the enlarged view of the thermal hysteresis of the FC and FCW curves in the shadow area.

To further illuminate the La-doping effect on the exchange anisotropy, isothermal magnetization measurements were performed at different temperatures in the thermal-hysteresis interval, as shown in Fig. 2e and f (full curves seen in Supplementary Figure 3). The areas of the magnetic loops are reasonably larger than those of the parent EuFeAs$_2$. EB emerges in both the *ab* and *c* directions, with a longitudinal shift along the magnetization axis. The EB fields for different temperatures are summarized in Fig. 2d. The increased $H_{EB}$s from those of the parent EuFeAs$_2$, and the longitudinal shift of the loop support the enhancement of the magnetic anisotropy in Eu$_{0.79}$La$_{0.21}$FeAs$_2$. A sign reversal of EB occurs below $T_M^{Eu}$ for **H** // **c**, while, below $T_{SR}$ for **H** // *ab*.

It is worth mentioning that higher magnetic field are needed to reverse the partially frozen moment at a lower temperature. Hence, lower magnetic field only results in *M-H* curves with loop area close to zero at a low temperature, shown in Supplementary Figure 3. The almost linear *M-H* curve shows an upward shift with bias in the field direction, which is similar to that observed in (Sm,La)FeO$_3$[33]. The upward shift can be ascribed to the pinning between the partially frozen magnetic moment and the reversible magnetic moment.

Briefly, La doping greatly affects the competing balance between the ferromagnetic and AF interactions of the $Eu^{2+}$ sublattice, and enhances the magnetic anisotropy in $Eu_{1-x}La_xFeAs_2$.

**Interplay of the anisotropic magnetism and SC.** To explore the interplay of the anisotropic magnetism and SC, an overdoped sample with a sharper superconducting transition is requisite. Here, a series of $Eu_{1-x}La_xFeAs_2$ ($x$ = 0.2, 0.25, and 0.3) polycrystalline samples are prepared and investigated. The chemical phase and quality of the samples are examined by powder X-ray diffraction (PXRD), detailed in Supplementary Figure 6. For these polycrystalline samples, we use the nominal doping levels to represent the La-doping contents. From the resistivity curves depicted in Fig. 3a, the anomaly related to the structural and Fe-AF transitions remains in these overdoped samples. As magnified in the inset of Fig. 3a, a sharp superconducting transition at $T_c \sim 11$ K is realized for $x$ = 0.2, which is suppressed with further doping. Zero resistivity is realized below $T_{zero} \sim 8.5$ K for $Eu_{0.8}La_{0.2}FeAs_2$. Hence, $Eu_{0.8}La_{0.2}FeAs_2$ is the expected superconducting compound.

Temperature dependences of magnetization were measured for these overdoped samples, as depicted in Fig. 3b. Eu-related ferromagnetic transition occurs in all the samples, different from the AF behavior of the compounds with $x \leq 0.15$[11]. The magnetization increases from $x$ = 0.2 to 0.25, indicating an enhancement of the ferromagnetic interaction. Then, it is suppressed by further La doping, implying an excessive magnetic dilution. Meanwhile, the SR-like behavior gradually disappears in these overdoped samples. A diamagnetic transition appears at 7 K for $Eu_{0.8}La_{0.2}FeAs_2$ in the ZFC process, with a superconducting volume fraction estimated to be 0.15–0.2 at 2.5 K. The broad superconducting transition for $Eu_{0.75}La_{0.25}FeAs_2$ in the $R$-$T$ curve disappears in the $M$-$T$ curve, which probably originates in the slight inhomogeneity nature for the polycrystalline sample and/or the filamentary SC.

ZFC isothermal magnetization was studied at 2.5 K for $Eu_{0.8}La_{0.2}FeAs_2$. The magnetic hysteresis loop obtained in a large field interval of -6 to 6 T is enlarged in Fig. 3c. The full and less-enlarged *M-H* curves can be seen in Supplementary Figure 4. The misalignment of the first quarter (0→6 T) and the last two quarters (-6→6 T) of the loop is due to the superposition of the superconducting and ferromagnetic loops. The superposed loop exhibits an EB behavior with $H_{EB}$ ~ -19 Oe, as seen from the comparison between the original and inverted curves.

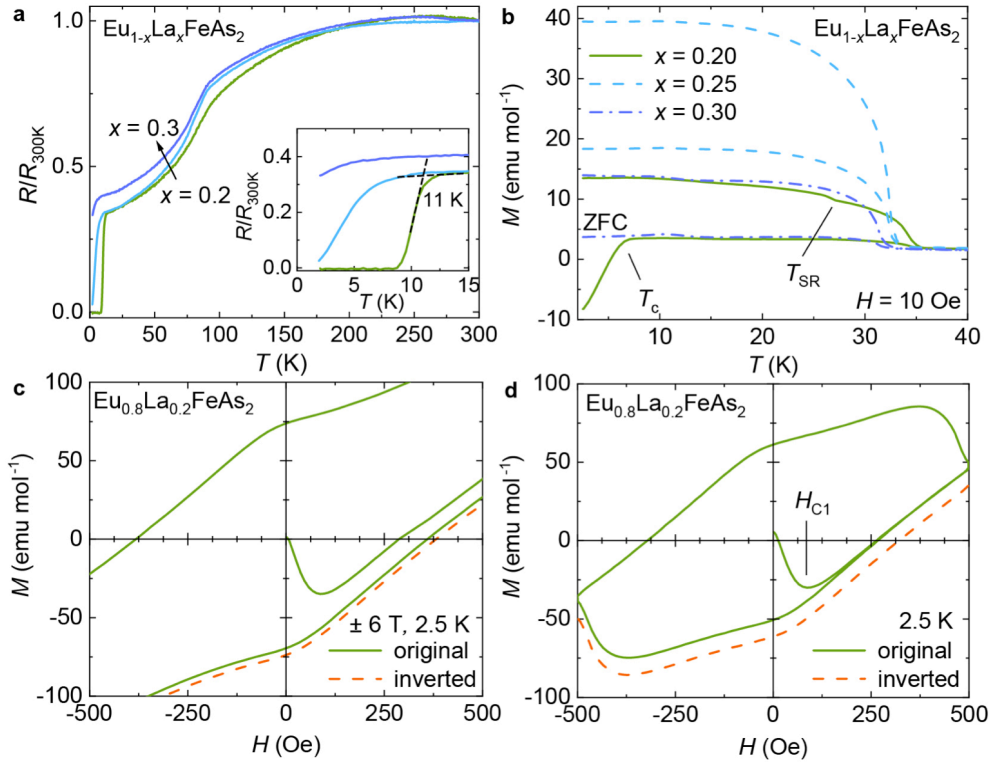

**Fig. 3 Physical properties of overdoped $Eu_{1-x}La_xFeAs_2$. a**, **b** Normalized resistivity and magnetization against temperature, respectively. **c**, **d** Magnetization against field for different field intervals of -6 to 6 T and -500 to 500 Oe, respectively, obtained at 2.5 K. The inset in **a** is a close view of the superconducting transition. Parts of the inverted $M$-$H$ curves, dashes in **c** and **d**, are for comparison with the original ones.

As mentioned above, lower magnetic field results in bias curves with loop area close to zero at lower temperatures. Thus, to eliminate the component of the ferromagnetic loop, the isothermal magnetization in a smaller field interval is studied, as shown in Fig. 3d. The first quarter and the fifth quarter of the loop coincide fast when passing the lower critical field $H_{C1}$, indicating that the superconducting loop is no longer superposed on the ferromagnetic loop. Meanwhile, an enhanced EB with $H_{EB}$ ~ -54 Oe is obtained.

In short, combining of the anisotropic magnetism and SC leads to an EB behavior of the superposed magnetic and superconducting loop for $Eu_{0.8}La_{0.2}FeAs_2$. This compound with interplay between SC and anisotropic magnetism may serve as a prototype for application exploration. Also, seeking for emergent physical phenomena from the interplay between multiple magnetic and superconducting orders are promising in this material.

**Nature of the Fe-AFM, and the FM/AFM/SC coexisting state.** To understand the nature of the robust Fe-AFM, we performed a density functional theory (DFT) calculation on the band structure of $EuFeAs_2$. Without considering the magnetic order of $Eu^{2+}$, seen in Fig. 4a, the band structures near the Fermi level are mainly attributed to the Fe-3d orbitals, of which the $t_{2g}$ orbitals contribute to the hole pockets at the Γ point and the electron pockets at the M point, similar to the band structure of LaFeAsO[34]. Following the magnetic structure of $EuFe_2As_2$[35], an assumed A-type AF order of $Eu^{2+}$ was considered in calculation, as seen in Fig. 4b. The band structures near the Fermi level barely changes, and the Eu-4f orbitals are below the Fermi level. Fig. 4c shows the Fermi surface (FS) of $EuFeAs_2$ with Fermi level $E_f$ = 0 lying at the charge neutral point. Similar to that in the Ca112 system[36], a reasonable FS nesting exists between the electron pockets at the M site and the hole pockets at the Γ point, suggesting the appropriate origin of the SDW-type AFM in EuFeAs$_2$. In order to further explore the influence of the electron doping, we artificially raise the Fermi level to examine the changes of the FS. The FSs with Fermi level $E_f$ = 30 and 50 meV (corresponding to 0.07 and 0.12 electron doping per Fe) are displayed in Fig. 4d and e, respectively, where the FS nesting is gradually weakened by electron doping but always exists.

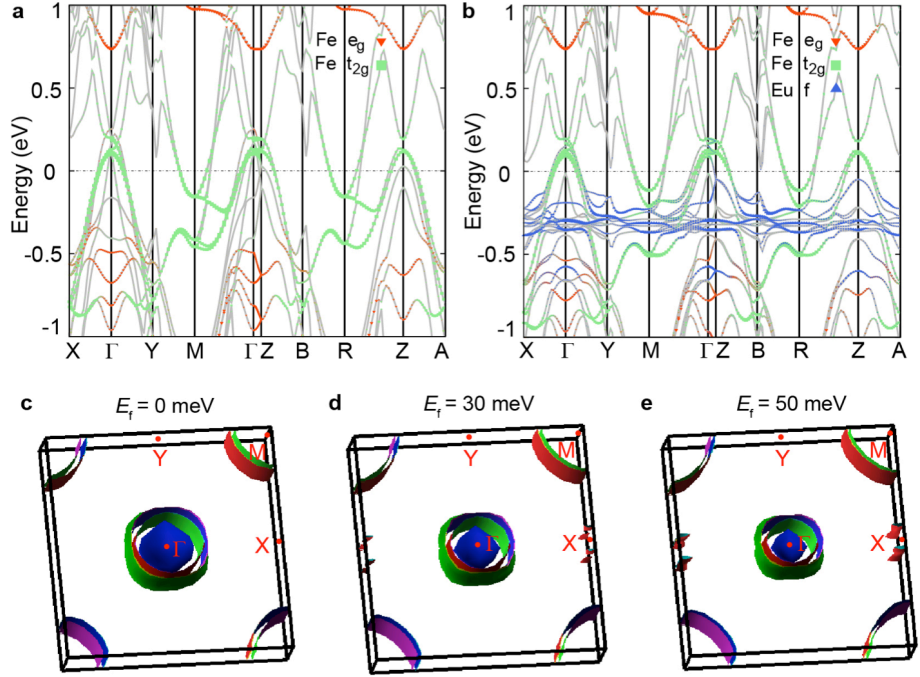

**Fig. 4 DFT calculations on $EuFeAs_2$. a, b** Band structures without and with A-type AF order at the $Eu^{2+}$ site, respectively, where the areas of the symbols represent the weights of the Fe d- and Eu f-orbitals. **c-e** Fermi surfaces with corresponding Fermi level $E_f$ = 0, 30, and 50 meV, respectively.

The SDW-type Fe-AFM of (Eu,La)FeAs$_2$ in the underdoping region can be explained by the FS nesting, despite the La- doping induced structural transformation[11,12], as discussed in Supplementary Figure 10. Whereas the FS nesting is gradually destroyed with electron doping exceeding 0.12 (Supplementary Figure 11). A dual itinerant and localized nature is proposed for the Fe-AFM in Ca$_{0.73}$La$_{0.27}$FeAs$_2$[36] and other iron-based systems[37]. Given that the ordered magnetic moment of Fe$^{2+}$ in EuFeAs$_2$ (0.78 $\mu_B$)[13] is relatively larger than many other iron-pnictide parents[38-40], we consider that the Fe-AFM in this Eu112 system is also dual-natured. Thus, with the FS nesting in (Eu,La)FeAs$_2$ weakened by La doping, the Fe-AFM in the overdoped region is probably contributed increasingly by the local superexchange interaction.

To further reveal the nature of the Fe-AFM in the overdoped area, as well as to check if it survives in the superconducting state, we performed a $^{57}$Fe Mössbauer spectroscopy investigation on the superconducting Eu$_{0.8}$La$_{0.2}$FeAs$_2$ polycrystalline sample. The fit of the $^{57}$Fe Mössbauer spectrum obtained at 300 K, detailed in Supplementary Figure 12, reveals a nearly single iron-containing phase. The fitted isomer shift (IS) and quadrupole splitting (QS) are 0.432(1) mm s$^{-1}$ and 0.157(4) mm s$^{-1}$, respectively, which are close to the corresponding values for the parent EuFeAs$_2$ and the Ni-doped EuFe$_{0.97}$Ni$_{0.03}$As$_2$[13].

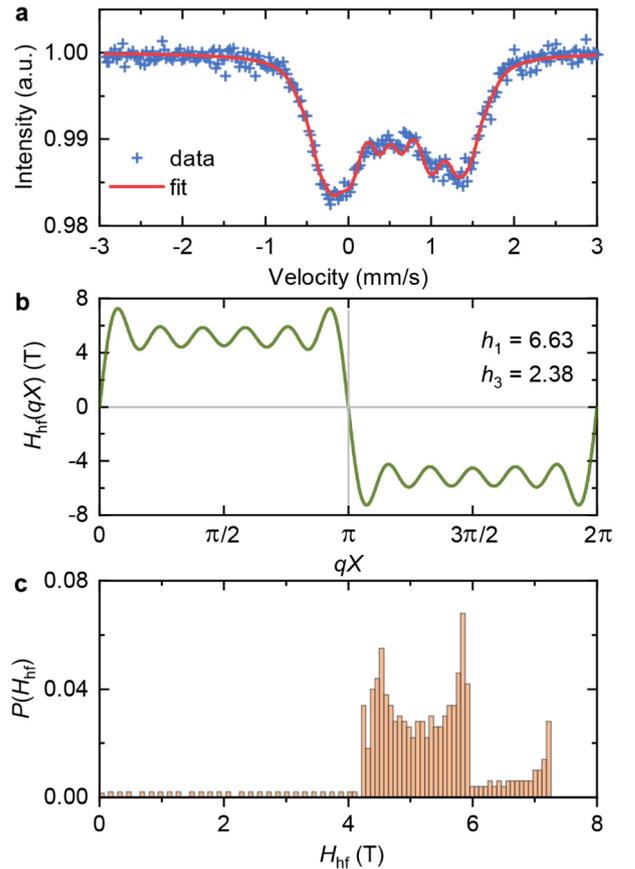

The spectrum collected at 6 K (< $T_c$), as shown in Fig. 5a, is similar to that of the undoped EuFeAs$_2$[13] in the form of a broadened, asymmetric, six-line Zeeman pattern, which can be explained by the distribution of hyperfine magnetic field due to the SDW-type AF order. To fit the spectrum of 6 K in the main text, we follow the procedure reported in Ref. [41]. In general, the hyperfine magnetic field of the spin-density-wave order can be expressed as

$$H(qX) = \sum_{n=1}^{N} h_{2n-1}\sin[(2n-1)qX], \quad (2)$$

where $h_{2n-1}$ denote the amplitudes of subsequent harmonics, $q$ stands for the wave number of the SDW, and $X$ denotes the relative position of the resonant nucleus along the propagation direction of the stationary SDW. The root-mean-square value of the hyperfine magnetic field $\sqrt{\langle H^2 \rangle}$ can be obtained as

$$\sqrt{\langle H^2 \rangle} = \sqrt{\tfrac{1}{2}\sum_{n=1}^{N} h_{2n-1}^2}, \quad (3)$$

which is proportional to the ordered magnetic moment $\mu_{Fe}$ carried by the Fe atoms. It is generally accepted that the magnetic moment is approximately proportional to the measured hyperfine magnetic field.

**Fig. 5** $^{57}$**Fe Mössbauer spectroscopy analysis on Eu$_{0.8}$La$_{0.2}$FeAs$_2$. a** The spectrum (blue crosses) obtained at 6 K and the fit (red solid line) with the SDW model detailed in the main text. **b** the SDW shape, and **c** the resulting hyperfine field distribution.

The obtained hyperfine parameters are listed in Supplementary Table 1, and the resulting SDW shape and the corresponding hyperfine field distribution are shown in Fig. 5b and c, respectively. The magnetic moment is determined to be 0.84(1) $\mu_B$ by using the same proportionality constant of $a$ = 63 kOe $\mu_B^{-1}$ as was used for the calculation of the magnetic moment of the parent compound $EuFeAs_2$[13]. The ordered magnetic moment is much larger than those of other iron-based superconducting samples with suppressed Fe-AFM[42-44]. Another interesting result is that the SDW shape is almost rectangular rather than quasi-triangular as found in most iron-based superconductors[13,42,43]. The rectangular SDW shape at a low temperature has been observed in some of the parent compounds with relatively large magnetic moment and less pronounced itinerant character[41]. Besides, the ratio of the third and first amplitudes $h_3/h_1$ ~ 0.36, which outclasses the range of $10^{-3}$–$10^{-2}$ expected from the itinerant-electron model[45-47], implies that the Fe-AFM in $Eu_{0.8}La_{0.2}FeAs_2$ cannot be accurately described merely by the itinerant picture. All these unusual Mössbauer spectroscopy results put our sample closer to the localized-AFM nature with the itinerant character of the magnetic order less prominent. Also, the magnetic moment is enhanced from that of the parent $EuFeAs_2$[13], which is in agreement with the increasing prominence of the local superexchange interaction suggested by the DFT calculation.

On the other hand, the Mössbauer spectrum obtained at 6 K manifests a microscopic coexistence of the Fe-AFM and SC. Given the relatively small superconducting volume fraction of $Eu_{0.8}La_{0.2}FeAs_2$, the robust Fe-AFM remains in the superconducting state probably with a cost of suppression on SC. Anyhow, $Eu_{0.8}La_{0.2}FeAs_2$ exhibits a microscopic coexistence of Eu-FM, Fe-AFM, and SC at low temperatures, similar to the Co-doped $EuFe_2As_2$ system[30,44].

Finally, combining the results above and the data we previously reported on the lightly-doped compounds[11], a La-doping electronic phase diagram on structure, magnetism, and SC for $Eu_{1-x}La_xFeAs_2$ is assembled in Fig. 6. All the values of the transition temperatures included in the phase diagram are listed in Supplementary Table 2. The La-doping-induced structural transformation occurs around $x$ ~ 0.05–0.1 (detailed in Supplementary Figure 7), which barely impacts the property evolution. The structural and Fe-AF transition temperatures are obtained from the derivation of the $R$-$T$ curves, see Supplementary Figure 9. The Fe-AF transition temperature of $Eu_{0.8}La_{0.2}FeAs_2$ obtained from the Mössbauer spectroscopy investigation (Supplementary Figure 13) is included for comparison, which manifests the reliability of the Fe-AF transition temperatures extracted from the $R$-$T$ data. Both the structural and Fe-AF transitions are slightly suppressed by La doping, but robustly remain. The slight suppression of the Fe-AFM by La doping in (Eu,La)FeAs$_2$ is likely due to the weakened FS nesting, which is contrary to the overdoped (Ca,La)FeAs$_2$ with stronger FS nesting and doping-enhanced Fe-AFM[10,36]. Consequently, the Fe-AFM phase with doping-adjustable dual nature is unusually adjacent to the whole superconducting dome. The robustness of the Fe-AFM is universal for electron doping in the Eu site, see the phase diagram of (Eu,Pr)FeAs$_2$ in Supplementary Figure 8. On the other hand, the Eu$^{2+}$ magnetic moments in EuFeAs$_2$ start to order below 45 K with a weak moment canting, leading to a coexistence of the Fe$^{2+}$ and Eu$^{2+}$ magnetic orders. The moment canting of the Eu$^{2+}$ sublattice is tunable by La doping, with the AF transition temperature suppressed with doping level increasing in the underdoped region. Ferromagnetism originating from the canted AF order of the Eu$^{2+}$ sublattice is realized for $x \geq 0.2$, with a higher ordering temperature than that of the AF transition temperature for $x = 0.15$, indicating the domination of the ferromagnetic interaction. With temperature further dropping, a superconducting dome is obtained by La doping. Under the dome, the superconducting order coexists with the Fe- and Eu- magnetic orders.

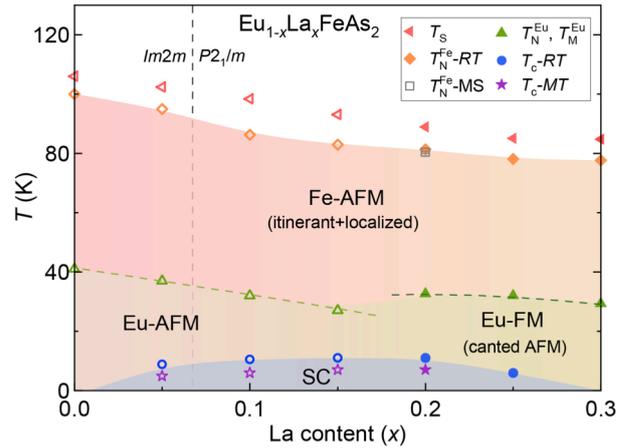

**Fig. 6 Electronic phase diagram of $Eu_{1-x}La_xFeAs_2$.** The structural, Fe-AF, Eu-related magnetic, and superconducting transition temperatures as functions of the nominal La doping content $x$ for the polycrystalline samples. The open symbols for $x \leq 0.15$ represent the data extracted from our previous work[11], and the solid symbols for $x \geq 0.15$ the data obtained in the present work. The structural transition temperatures ($T_S$), the Fe-AF transition temperatures ($T_N^{Fe}$-$RT$), and the superconducting transition temperatures ($T_c$-$RT$) are obtained from the transport measurements. The Fe-AF transition temperature ($T_N^{Fe}$-$MS$) for $x$ = 0.2 is obtained from the Mössbauer spectroscopy investigation, where the error bar represents the s.e.m. The Eu-magnetic transition temperatures ($T_N^{Eu}$ and $T_M^{Eu}$) and the diamagnetic transition temperatures ($T_c$-$MT$) are obtained from the magnetization measurements.

In summary, we systematically investigated the electrical and magnetic properties of the 112-type (Eu,La)FeAs$_2$. Due to the magnetic anisotropy, various exceptional magnetic phenomena are discovered in the parent EuFeAs$_2$. Nonmagnetic La substitution modifies the balance of the ferromagnetic-AF competition and enhances the magnetic anisotropy. Several related physical phenomena are further revealed, including the EB effect of the superposed ferromagnetic/superconducting loop; the robustness of the Fe-AFM with doping-adjustable dominance of the dual itinerant and localized nature; and the coexisting state of Eu-FM, Fe-AFM, and SC. We call for further theoretical explanations for the SDW-associated magnetic exchange anisotropy. The incorporation of superconducting electrons and anisotropic spin states may trigger explorations of applications in electronics and spintronics, for example, in the cross-control field. Experimental investigations of the underlying physical phenomena in the FM/AFM/SC coexisting state are promising, given the coexistence of the multiple orders and the strong couplings. Most importantly, SC adjacent to the Fe-AFM with doping-adjustable itinerant/localized characters may host different threads to the nature of high-temperature SC in different doping regions.

**METHODS**

Sample preparation

Single crystals of EuFeAs$_2$ and Eu$_{0.79}$La$_{0.21}$FeAs$_2$ were grown from a CsCl flux. A mixture of elementary Eu/La, Fe, and As in ratio of 1 : 1 : 4 (or 2 : 1 : 6) with 10- to 20-fold of dehydrated CsCl was sealed in a vacuum quartz tube, heated slowly to 800 °C, and held for 2 weeks before quenching. Polycrystalline Eu$_{1-x}$La$_x$FeAs$_2$ ($x = 0.2, 0.25$, and $0.3$) samples were synthesized following our previous work[11]. The reaction temperature in the last step was modulated to 850 °C to improve the La-doping homogeneity in the overdoped samples.

Phase and Property characterization

The SXRD experiments were carried out on a Single-crystal X-ray Diffractometer (Bruker). The PXRD patterns were collected on a Powder X-ray Diffractometer (PAN-analytical). The EDXS experiment was performed using a Scanning Electron Microscope (SEM) equipped with an Energy Dispersive X-ray Spectrometer (ZEISS). Electrical transport, heat capacity, and magnetic measurements were conducted on a Physical Property Measurement System (PPMS) and a Magnetic Property Measurement System (MPMS) (Quantum Design).

Transmission $^{57}$Fe Mössbauer spectra were recorded by using a conventional spectrometer working in constant acceleration mode. A 50 mCi of $^{57}$Co embedded in a Rh matrix moving at room temperature was used as the γ-ray source. The absorber was prepared with a surface density of ~8 mg cm$^{-2}$ natural iron. The drive velocity was calibrated with sodium nitroprusside at room temperature and all the isomer shifts quoted in this work are relative to that of the α-Fe.

Theoretical calculations

Theoretical calculations were performed using the DFT as implemented in the Vienna *ab initio* simulation package (VASP) code[48-50]. The generalized-gradient approximation (GGA) for the exchange correlation functional was used. The cutoff energy was set to be 400 eV for expanding the wave functions into plane-wave basis. In the calculation, the BZ was sampled in the *k* space within Monkhorst-Pack scheme[51].

## DATA AVAILABILITY

The data that support the findings of this study are available from the corresponding authors upon reasonable request.

## ACKNOWLEDGEMENTS

We would like to thank Xian-Xin Wu and Lv-Kuan Zou for useful discussions, and thank Fan Cui for the help on the figures of the FSs. The work was supported by the National Natural Science Foundation of China (11904414, 11774402, 11704167); National Key Research and Development Program of China (2019YFA0705702); the National Key Research Program of China (2018YFA0704200, 2016YFA0300301); and the Fundamental Research Funds for the Central Universities (2021qntd27).

# Supplementary Information for
# Coexistence of ferromagnetism, antiferromagnetism, and superconductivity in magnetically anisotropic (Eu,La)FeAs$_2$

Jia Yu[1*], Congcong Le[2], Zhiwei Li,[3] Lisi Li,[1] Tong Liu[4,5], Zengjia Liu,[1] Bo Zhang,[3] Bing Shen[1], Binbin Ruan[4,5], Zhi'an Ren[4,5*], Meng Wang[1*]

[1] Center for Neutron Science and Technology, School of Physics, Sun Yat-sen University, Guangzhou 510275, China

[2] Max Planck Institute for Chemical Physics of Solids, 01187 Dresden, Germany

[3] Key Lab for Magnetism and Magnetic Materials of the Ministry of Education, Lanzhou University, Lanzhou 730000, China

[4] Beijing National Laboratory for Condensed Matter Physics, Institute of Physics, Chinese Academy of Sciences, Beijing 100190, China

[5] School of Physical Sciences, University of Chinese Academy of Sciences, Beijing 100049, China

*Corresponding authors. E-mails: jyu_work@163.com; wangmeng5@mail.sysu.edu.cn; renzhian@iphy.ac.cn

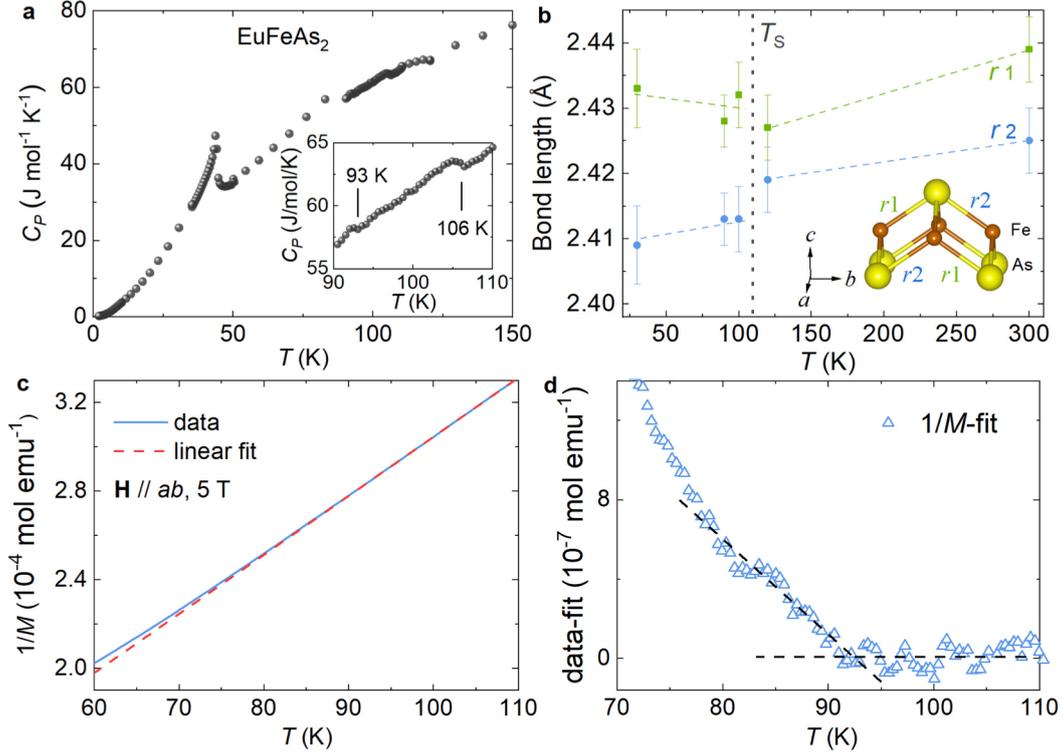

**Supplementary Figure 1: Phase transitions of EuFeAs$_2$.** Temperature dependences of **a** heat capacity, **b** length evolution of the FeAs bonds, and **c** reciprocal of the magnetization with magnetic field of 5 T parallel to the *ab* plane and the corresponding linear fit. **d** displays the reciprocal of the magnetization after subtracting the fitting line, where the dash lines are guides for the eye.

Heat capacity measurement from 150 K to 2 K is performed on single crystalline EuFeAs$_2$ to confirm the phase transitions, as depicted in Supplementary Figure 1a. Two associated slight transitions around 106 K and 93 K, and one significant transition around 45 K are observed, which agrees with the previous electrical transport and magnetization results[1,2].

The significant transition around 45 K is in consistent with the canceling of the large moment of Eu$^{2+}$ from the magnetic measurement. Focusing on the two associated transitions around 100 K, we conducted an SXRD study on EuFeAs$_2$ at different temperatures. The crystal structures for different temperatures share the same space group *Im2m*. We carefully checked the configuration of the FeAs layer, and found that a distortion occurs between 120 K and 100 K. The lengths of the Fe-As bonds in the *bc* plane undergoes a clear change, as displayed in Supplementary Figure 1b. Thus, the phase transition around 106 K is associated with the FeAs-layer-related structural distortion.

The temperature dependence of magnetization was measured under a magnetic field of 5 T parallel to the *ab* plane. The reciprocal of the magnetization is scrutinized around 100 K, see Supplementary Figure 1c. According to the Curie-Weiss law, the 1/*M-T* curve should be linear with a slope related to the effective moment in the paramagnetic region. We found that the experimental data deviate from the linear fitting at low temperatures. After subtracting the fit, the data exhibit clearly a reduce of slope at 93 K, see Supplementary Figure 1d, which is a solid evidence of the moment canceling. Thus, combining the heat capacity and magnetization results, as well as the Mössbauer spectroscopy investigation on the polycrystalline EuFeAs$_2$[3], we conclude that the Fe-AF transition occurs at $T_N^{Fe} \sim 93$ K in single crystalline EuFeAs$_2$.

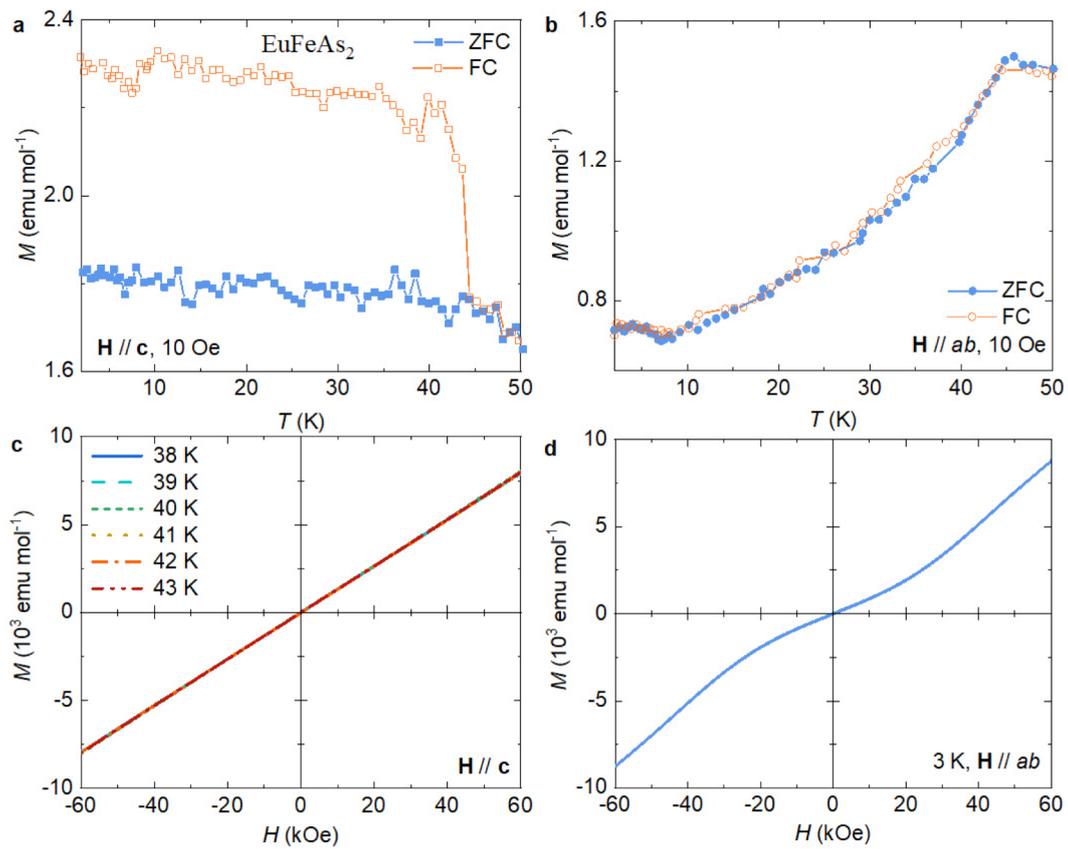

**Supplementary Figure 2: Magnetic properties of EuFeAs$_2$. a**, **b** Magnetization against temperature under a magnetic field of 10 Oe parallel to different directions; and **c**, **d** The full *M-H* curves of those enlarged in Fig. 1 in the main text for field parallel to different directions, for single crystalline EuFeAs$_2$.

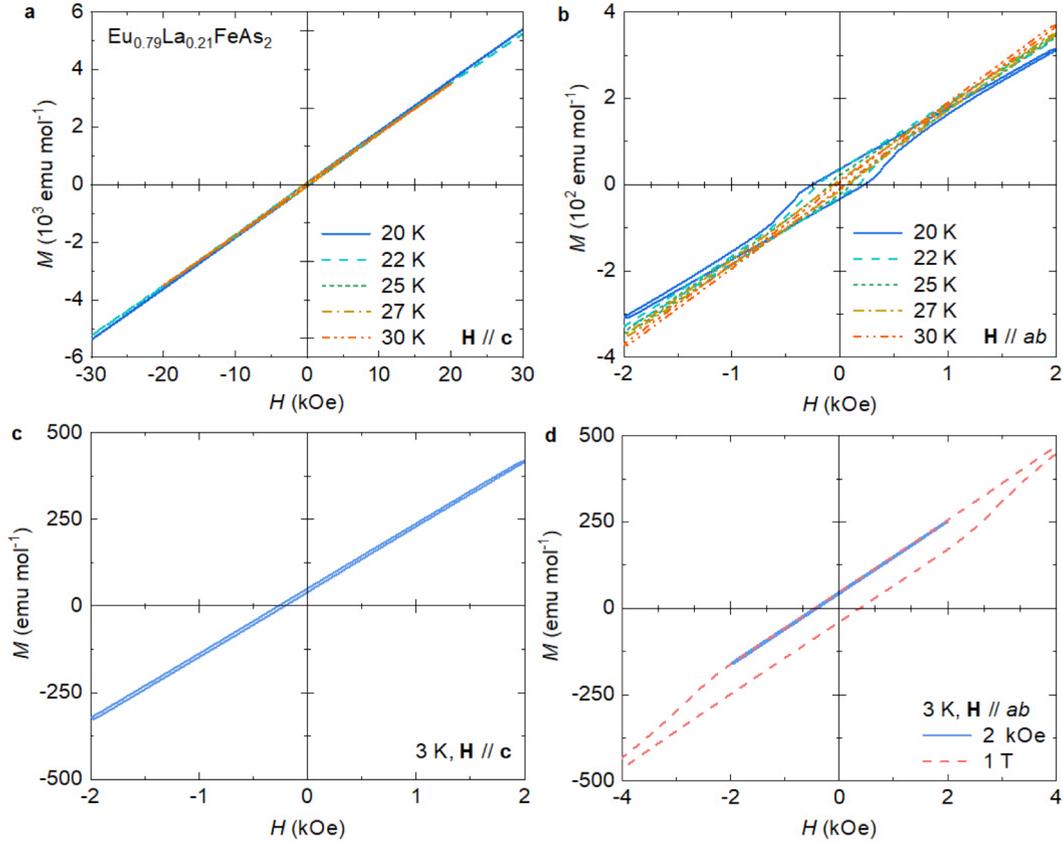

**Supplementary Figure 3: Magnetic properties of Eu$_{0.79}$La$_{0.21}$FeAs$_2$.** **a, b** The full *M-H* curves of those enlarged in Fig. 2 in the main text for field parallel to different directions at temperatures from 20 to 30 K; and **c, d** the *M-H* curves for field parallel to different directions obtained at 3 K in small field intervals, for single crystalline Eu$_{0.79}$La$_{0.21}$FeAs$_2$.

From the full *M-H* curves (of those enlarged in Fig. 2 in the main text) at higher temperatures in Supplementary Figure 3a and b, magnetic field of 2 kOe is sufficient to form a hysteresis loop for **H** // *ab* in single crystalline Eu$_{0.79}$La$_{0.21}$FeAs$_2$, however, much larger magnetic field is requested in the *c* direction. With temperature dropping, larger magnetic field is needed to form a hysteresis loop. A magnetic field of 2 kOe cannot reverse the spin at 3 K, and results in a bias loop with area close to zero in both directions, seen in Supplementary Figure 3c and d. With field increased to 1 T, spin is successfully reversed, resulting in a regular hysteresis loop at 3 K for **H** // *ab*.

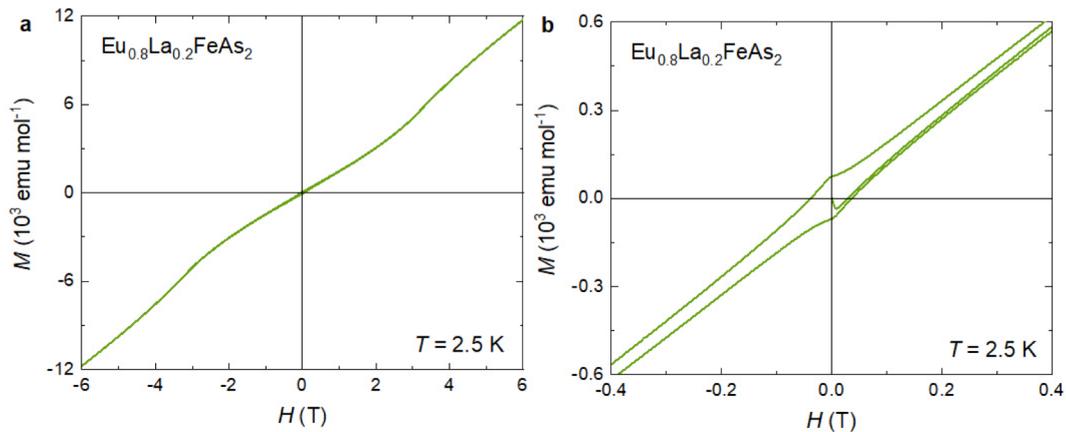

**Supplementary Figure 4: Magnetic properties of Eu$_{0.8}$La$_{0.2}$FeAs$_2$.** **a** The full and **b** less-enlarged *M-H* curves of the more-enlarged one in Fig. 3 in the main text for the polycrystalline Eu$_{0.8}$La$_{0.2}$FeAs$_2$.

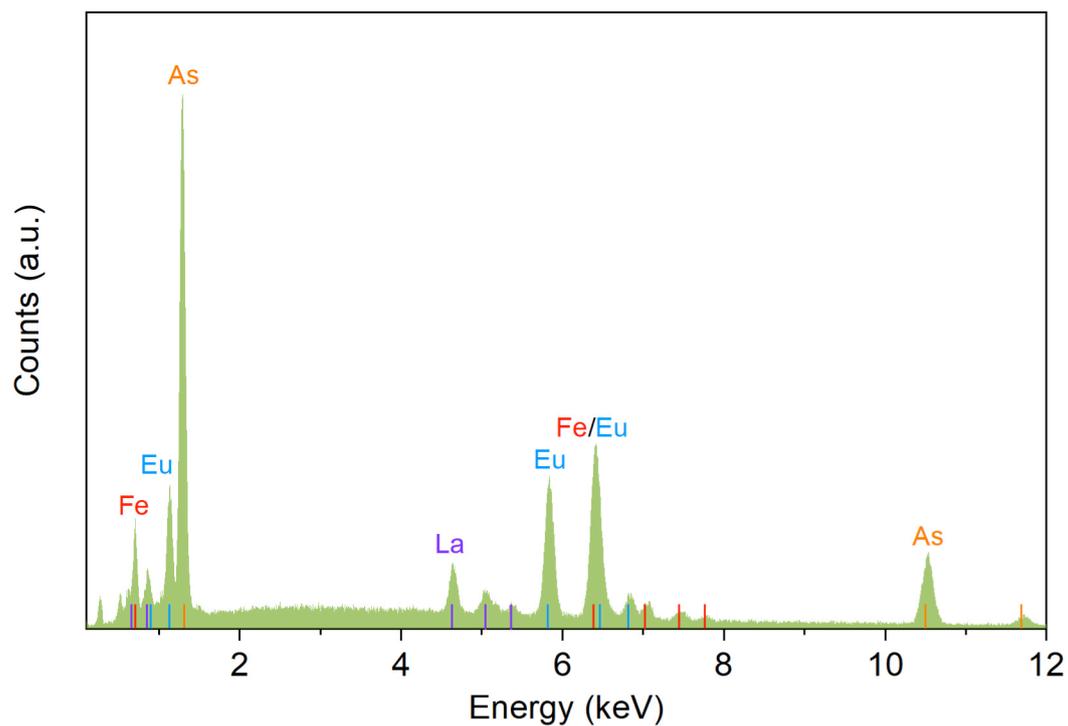

**Supplementary Figure 5:** EDXS spectrum for the $Eu_{0.79}La_{0.21}FeAs_2$ single crystal. The indexing bars indicate the peak positions of Eu, La, Fe, and As.

The EDXS spectrum for the overdoped $(Eu,La)FeAs_2$ single crystal (investigated in the main text), seen in Supplementary Figure 5, is mainly contributed by the elements of Eu, La, Fe, and As. The obtained molar ratio of Eu : La : Fe : As = 19.27 : 5.08 : 24.80 : 50.85, resulting in the chemical formula of $Eu_{0.79}La_{0.21}FeAs_2$.

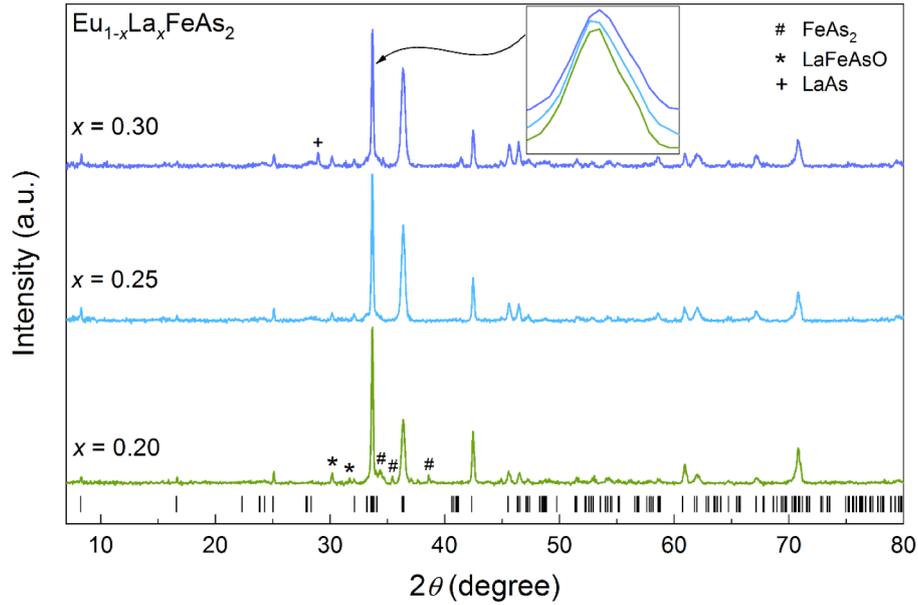

**Supplementary Figure 6:** PXRD patterns for La-overdoped $Eu_{1-x}La_xFeAs_2$ ($x$ = 0.2, 0.25, and 0.3), comparing with the peak indexes calculated from the space groups of $P2_1/m$[1] at the bottom.

The PXRD patterns for the overdoped polycrystalline samples studied in the main text were indexed with space group $P2_1/m$[1], seen in Supplementary Figure 6. Only minor peaks with weak intensity for trace impurities present. The peak shifting, as enlarged in the inset, indicates a structural shrinkage due to the substitution of larger $Eu^{2+}$ by smaller $La^{3+}$. The appearance of LaAs for $x$ = 0.3 implies that the limit of the La solubility in $EuFeAs_2$ is near 30%. Generally, the trace impurities and the clear peak shifting imply a successful doping of these overdoped polycrystalline samples. Note that the actual La-doping level for polycrystalline samples is a little smaller than the nominal ones due to the La-related residuals, which is the reason why the polycrystalline $Eu_{0.8}La_{0.2}FeAs_2$ and the $Eu_{0.79}La_{0.21}FeAs_2$ single crystal show different superconducting behaviors (in the main text).

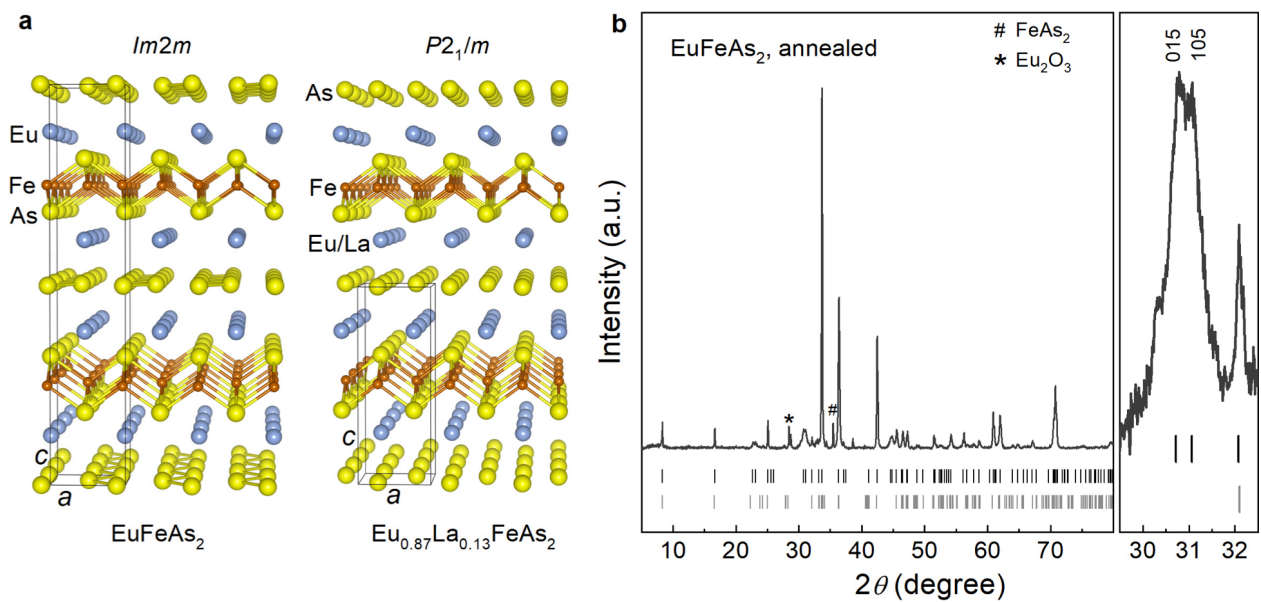

**Supplementary Figure 7: Structural transformation of $Eu_{1-x}La_xFeAs_2$.** **a** Crystal structure of orthorhombic $EuFeAs_2$ and monoclinic $Eu_{0.87}La_{0.13}FeAs_2$ from the SXRD experiments[1,2]. **b** PXRD pattern for annealed polycrystalline $EuFeAs_2$ with peak indexes calculated from the space groups of $Imm2$ (up, black) and $P2_1/m$ (down, grey) at the bottom, and magnified pattern around 31° to the right.

Our previous studies revealed a monoclinic structure with space groups of $P2_1/m$ for $Eu_{0.87}La_{0.13}FeAs_2$ by SXRD analysis, and the PXRD patterns of the polycrystalline $Eu_{1-x}La_xFeAs_2$ ($x$ = 0–0.15) were indexed based on this monoclinic structure[1]. Later, we found that the structure of the undoped $EuFeAs_2$ belongs to space group $Imm2$[2] (same to $Im2m$ in present work, with a change of coordinates), shown in Supplementary Figure 7a. It is necessary to figure out the structural transformation induced by La doping. The PXRD patterns for the polycrystalline (Eu,La)$FeAs_2$ samples are not of high quality for structural refinement due to the naturally poor crystallinity. Thus, we compared the PXRD pattern of an annealed $EuFeAs_2$ sample and the peak positions calculated from both space groups, as seen in Supplementary Figure 7b. Though the quality of the pattern is still not high enough, two peaks, around $2\theta \sim 31°$ can be indexed with space group $Im2m$ only, agreeing with the SXRD result. By reexamining the patterns of the as-grown $Eu_{1-x}La_xFeAs_2$ samples in our previous work[1], the hump around 31° for $x$ = 0 and 0.05 corresponds with the (0 1 5) and (1 0 5) peaks of $Im2m$, which disappears for $x$ = 0.1 and 0.15. Thus, the La-doping induced structural transformation occurs around doping level $x$ between 0.05 and 0.1.

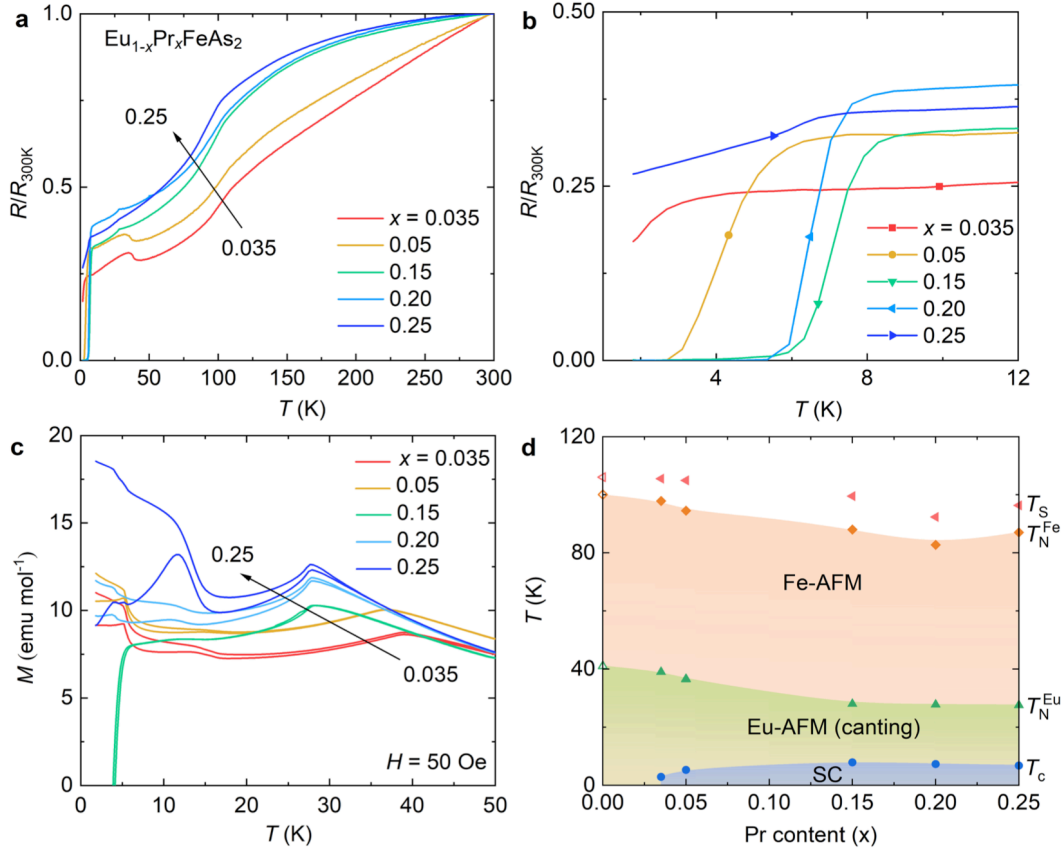

**Supplementary Figure 8: Physical properties of $Eu_{1-x}Pr_xFeAs_2$.** The electrical and magnetic properties of $Eu_{1-x}Pr_xFeAs_2$: **a** temperature dependences of resistivity, which are enlarged around the superconducting transition area in **b**; **c** magnetization against temperature; and **d** doping phase diagram, where the open symbols for the parent phase represent the data extracted from our previous work[1]. The structural and Fe-AF transition temperatures ($T_S$ and $T_N^{Fe}$) are obtained from the first derivation of the $R$-$T$ curves, see Supplementary Figure 9b. The Eu-AF transition temperatures ($T_N^{Eu}$) are obtained from the $M$-$T$ curves. The superconducting transition temperatures ($T_c$) are obtained from the $R$-$T$ curves. All the values in the phase diagram are listed in Supplementary Table 2.

A series of Pr-doped $Eu_{1-x}Pr_xFeAs_2$ ($x = 0.035, 0.05, 0.15, 0.2,$ and $0.25$) polycrystalline samples were synthesized for the comparison with the doping effect of $Eu_{1-x}La_xFeAs_2$. The electrical properties for $Eu_{1-x}Pr_xFeAs_2$, as seen in Supplementary Figure 8a and b, are similar to that of $Eu_{1-x}La_xFeAs_2$. However, the magnetic properties, seen in Supplementary Figure 8c, are different. The mild moment canting behavior for all the samples indicate that the net moment in $Eu_{1-x}Pr_xFeAs_2$ is smaller than that in $Eu_{1-x}La_xFeAs_2$, which implies a difference originating from the different degrees of magnetic dilution. The phase diagram, Supplementary Figure 8d, exhibits a similar relationship between the robust Fe-AFM and SC. Moreover, an enhancement of the Fe-AFM appears from $x = 0.2$ to $0.25$. Note, the open symbols in the phase diagram for $x = 0$ represent the data extracted from our previous work[1].

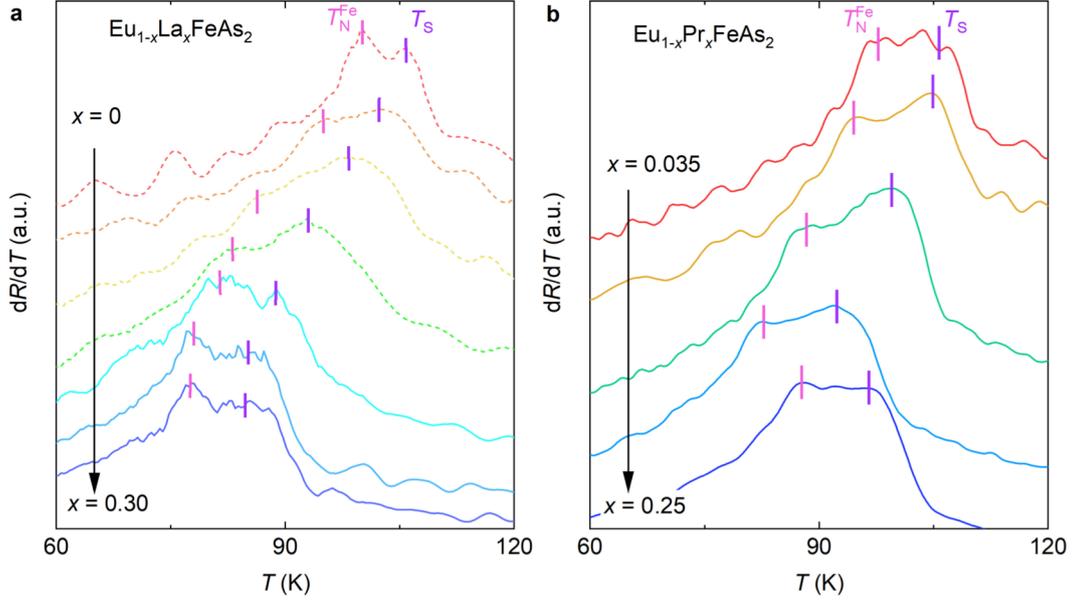

**Supplementary Figure 9: Determination of the structural and Fe-AF transition temperatures.** The first derivation of the $R$-$T$ curves for the polycrystalline samples of **a** $Eu_{1-x}La_xFeAs_2$ and **b** $Eu_{1-x}Pr_xFeAs_2$ in a temperature interval around the structural and Fe-AF transitions. The dash lines in **a** are obtained based on the $R$-$T$ curves in our previous work[1].

The structural phase transition temperatures ($T_S$) and the Fe-AF phase transition temperatures ($T_N^{Fe}$) for the polycrystalline $Eu_{1-x}La_xFeAs_2$ and $Eu_{1-x}Pr_xFeAs_2$ samples included in the phase diagrams are obtained from the first derivation of the corresponding $R$-$T$ curves, as shown in Supplementary Figure 9a and b, respectively. The dash lines in Supplementary Figure 9a are obtained based on the $R$-$T$ curves extracted from our previous work[1]. The structural and Fe-AF transition temperatures can be roughly located from the peaks related to the transitions. Both the structural and Fe-AF transitions show robustness against La or Pr doping even in the overdoped region.

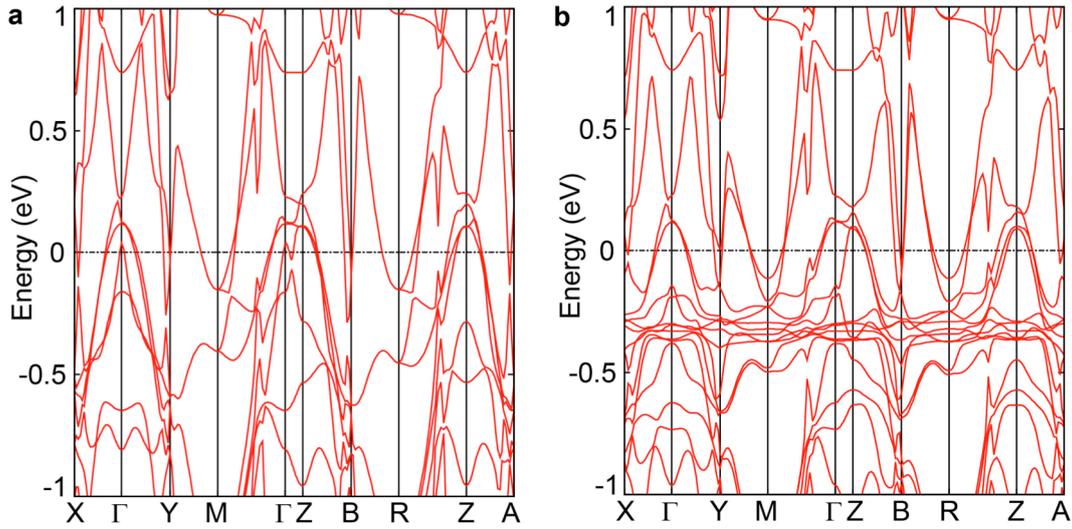

**Supplementary Figure 10:** Calculated band structures of EuFeAs$_2$ **a** without and **b** with assumed A-type AF order at the Eu$^{2+}$ site from DFT calculation, where we use the lattice parameters of Eu$_{0.87}$La$_{0.13}$FeAs$_2$[1] with monoclinic $P2_1/m$ structure.

To support the procedure of using the FS nesting to explain the evolution of the Fe-AFM with the existence of the La-doping-induced structural transformation, we calculated the band structures of undoped EuFeAs$_2$ using the $P2_1/m$ structure. Comparing to the band structures calculated based on the $Im2m$ structure in the main text, the bands attributed to the Fe-3d orbitals near the fermi level based on the $P2_1/m$ structure[1] barely changes, as shown in Supplementary Figure 10, suggesting that we can artificially raise the Fermi level in the band structures of parent EuFeAs$_2$ to simulate the band structures of the La-doped Eu$_{1-x}$La$_x$FeAs$_2$, despite the structural transformation.

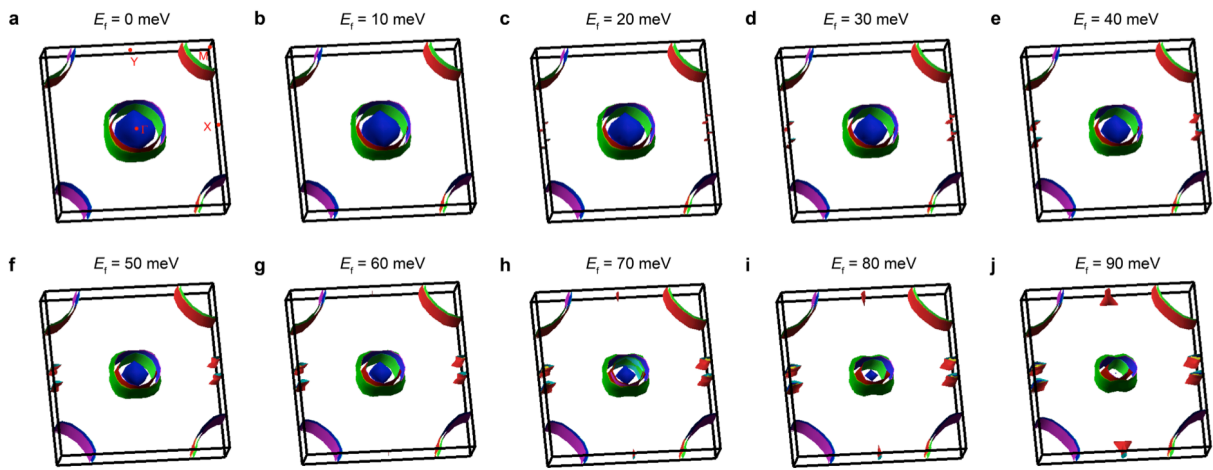

**Supplementary Figure 11: a-j** The evolution of the FSs of EuFeAs$_2$ with Fermi level from $E_f = 0$ to 90 meV.

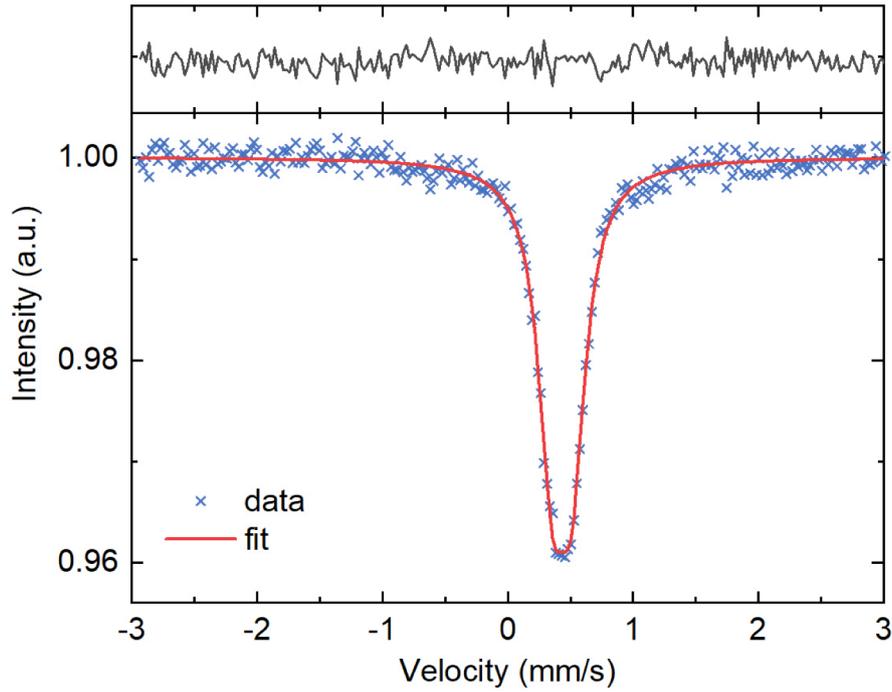

**Supplementary Figure 12:** $^{57}$Fe Mössbauer spectrum of Eu$_{0.8}$La$_{0.2}$FeAs$_2$ collected at 300 K, with transmission integral fit using one doublet (red solid line). The residual of the fit is shown in the upper panel with a same scale.

The $^{57}$Fe Mössbauer spectrum obtained at 300 K, seen in Supplementary Figure 12, can be reasonably fitted with only one doublet indicating that the iron-containing impurity phases should be less than the detection limit of ~2%, which is in consistent with the PXRD analysis in Supplementary Figure 6.

**Supplementary Table 1:** Mössbauer hyperfine parameters at 6 K determined from the fit described above.

| | |
|---|---|
| IS (mm s$^{-1}$) | 0.536(7) |
| QS (mm s$^{-1}$) | 0.05(1) |
| $h_{2n-1}$ (T) | 6.63, 2.38, 1.59, 1.24, 1.02, 1.15 |
| $\sqrt{\langle H^2 \rangle}$ (T) | 5.3(2) |
| $\mu_{Fe}$ ($\mu_B$) | 0.84(1) |

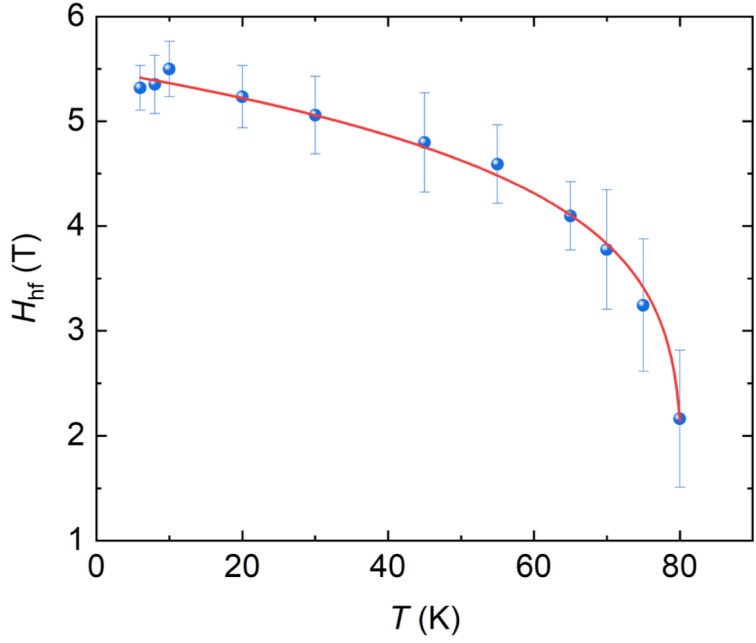

**Supplementary Figure 13:** Hyperfine magnetic fields (blue scatters) at $^{57}$Fe nuclei obtained from the Mössbauer spectrum at different temperatures, where the error bars represent the s.e.m. The red line is the power-law fitting of the $H_{hf}$ data.

$^{57}$Fe Mössbauer spectroscopy investigations were also carried out at different temperatures in the range of 6-80 K. The spectrum at each temperature is fitted with the SDW model (not shown). The obtained hyperfine magnetic fields $H_{hf}$s at $^{57}$Fe nuclei are summarized in Supplementary Figure 13. The $H_{hf}$-$T$ data are fitted with the power law, yielding the hyperfine magnetic field $H_{hf}(0) \sim 5.49(4)$ T at 0 K, and the Fe-AF transition temperature $T_N^{Fe} \sim$ 80.4(4) K. A slight dropping of $H_{hf}$ can be observed below $T_c$, which we ascribe to the competition between SDW and SC.

**Supplementary Table 2:** The transition temperatures (Kelvin) included in the phase diagrams.

|  | Eu$_{1-x}$La$_x$FeAs$_2$ | | | | | | Eu$_{1-x}$Pr$_x$FeAs$_2$ | | | |
|---|---|---|---|---|---|---|---|---|---|---|
|  | $T_S$ | $T_N^{Fe}$ − RT | $T_N^{Fe}$ − MS | $T_N^{Eu}, T_M^{Eu}$ | $T_c$-RT | $T_c$-MT | $T_S$ | $T_N^{Fe}$ | $T_N^{Eu}$ | $T_c$ |
| $x = 0$ | 106 | 100 |  | 41 |  |  |  |  |  |  |
| 0.035 |  |  |  |  |  |  | 105.5 | 97.8 | 39 | 2.9 |
| 0.05 | 102.4 | 95 |  | 37 | 8.87 | 4.9 | 104.9 | 94.5 | 36.5 | 5.3 |
| 0.10 | 98.4 | 86.3 |  | 32 | 10.47 | 6 |  |  |  |  |
| 0.15 | 93.1 | 83 |  | 27 | 11 | 7 | 99.5 | 88 | 28 | 7.9 |
| 0.20 | 88.9 | 81.4 | 80.4(4) | 32.7 | 11 | 7 | 92.3 | 82.7 | 27.8 | 7.3 |
| 0.25 | 85.1 | 78.1 |  | 32.1 | 6 |  | 96.3 | 87 | 27.6 | 6.8 |
| 0.30 | 84.8 | 77.7 |  | 29.4 |  |  |  |  |  |  |